\documentclass[twocolumn,superscriptaddress,nofootinbib]{revtex4-1}
\usepackage{graphicx}
\usepackage{dcolumn}
\usepackage{bm}
\usepackage{subfigure}
\usepackage{mathrsfs}
\usepackage{multirow}
\usepackage{amsmath}
\usepackage{amsfonts}
\usepackage[makeroom]{cancel}

\usepackage[colorlinks,
linkcolor={red!60!black!},
anchorcolor={green!80!black!},
urlcolor={blue!80!black!},
citecolor={blue!50!black!}]{hyperref}
\usepackage[table]{xcolor}
\usepackage{tipa}

\newcommand{\G}[8]{G\left(\overline{#1#2},\overline{#3#4};\overline{#5#6},\overline{#7#8}\right)}
\newcommand{\Gz}[4]{G\left(\overline{0#1},\overline{0#2};\overline{0#3},\overline{0#4}\right)}

\begin{document}

\title{Twisted geometries are area-metric geometries}
\author{Bianca Dittrich} 
\email{bdittrich@perimeterinstitute.ca}

\affiliation{Perimeter Institute, 31 Caroline Street North, Waterloo, ON, N2L 2Y5, Canada
}

\author{José Padua-Arg\"uelles} 
\email{jpaduaarguelles@perimeterinstitute.ca}

\affiliation{Perimeter Institute, 31 Caroline Street North, Waterloo, ON, N2L 2Y5, Canada
}
\affiliation{Department of Physics and  Astronomy, University of Waterloo, 200 University Avenue West, Waterloo, ON, N2L 3G1, Canada}


\begin{abstract}

The quantum geometry arising in Loop Quantum Gravity has been known to semi-classically lead to generalizations of length-geometries. There have been several attempts to interpret these so called twisted geometries and understand their role and fate in the continuum limit of the spin foam approach to quantum gravity. In this paper we offer a new perspective on this issue by showing that the twisted geometry of a 4-simplex can be understood as arising from an area-metric (in contrast to the more particular length-metric).

Such equivalence allows us to define notions like signature, generalized triangle inequalities and parallel transport for twisted geometries (now understood in a 4-dimensional setting), exemplifying how it provides a new handle to understand them. Furthermore, it offers a new microscopic understanding of spin foam geometries which is notably supported by recent studies of the continuum effective dynamics of spin foams. 

\end{abstract}

\maketitle

\section{Introduction \label{sec:intro}}

\emph{Grosso modo}, the covariant Loop Quantum Gravity (LQG), or spin foam, approach to quantum gravity can be understood as coming in three main steps: One first discretizes geometry and quantizes that, then imposes (discrete) gravitational dynamics to finally take a refinement limit. The discretization step can be intuitively understood as approximating space-time using piece-wise flat geometries obtained by gluing convex polytopes together, with the simplest one being a simplex.
Although it arose as a conservative approach that does not assume \emph{a priori} additional degrees of freedom beyond those of general relativity, it actually does have an extended configuration space of four-dimensional geometries \cite{Dittrich:2008va,Dittrich:2008ar}.

Such a parametrization results from the fact that a quantum 4-simplex is understood as a gluing of five quantum 3-simplices (tetrahedra) \cite{Baez:1999tk}. Indeed, the Hilbert space associated to a 4-simplex has as a basis the set of spin network states of its boundary, which precisely represent the quantum geometry of its five constituent tetrahedra.\footnote{Equivalently, one can note that the EPRL/FK spin foam amplitude \cite{Engle:2007uq,Freidel:2007py} of a 4-simplex has fifteen labels. Or follow a canonical analysis similar to that of \cite{Dittrich:2008ar}.}

More precisely, as argued in \S\ref{ssec:QSG}, to each tetrahedron one can associate five quantum numbers, so that gluing them together would require us to match the 2-geometry of the ten triangles shared by pairs of tetrahedra. This would amount to implementing several constraints which are quantum equivalents of matching the triangles' lengths as seen from the different tetrahedra \cite{Dittrich:2008va}. However, this process is hindered by quantum uncertainty, that only allows us to implement a subset of the constraints exactly, \emph{e.g.} setting only the areas of the matched triangles to be equal.  The remaining constraints lead to an enlargement of the classical 4-simplex configuration space, which is parametrized by 10 numbers, to the quantum configuration space, which is parametrized by  $5\times5-10=15$ quantum numbers.

In this paper we are interested in the enlarged classical configuration space which these geometries quantize and its geometrical meaning, from a four dimensional perspective.\footnote{For a three dimensional point of view, more akin to thinking just on boundary triangulations, see \cite{Dittrich:2008ar,Freidel:2010aq,Freidel:2010bw,Haggard:2012pm,Dittrich:2010ey,Dittrich:2012rj,Freidel:2013bfa}.} As confirmed by analyzing an over-complete basis of coherent states for the boundary of a 4-simplex, it is one obtained by gluing classical tetrahedra together (whose geometry is fixed by six independent parameters —\emph{e.g.} the edge lengths), but only matching the areas of triangles shared by pairs so that their shapes may not match, thus it is parametrized by $6\times 5-10=20$ independent quantities\footnote{
The reader might wonder why one has 15 quantum numbers but 20 classical parameters describing the configuration space. The reason is that out of the 20 parameters one has 10 parameters, given by 3D dihedral angles, which form non-commuting pairs with respect to the Poisson brackets provided by the underlying phase space \cite{Dittrich:2008ar,Dittrich:2010ey}. (See also the discussion in  \S\ref{ssec:QSG}.) These 10 parameters result therefore only in 5 quantum numbers. }
\cite{Freidel:2013fia}. This is indeed larger than the number of parameters needed to fix the usual length geometry of a 4-simplex. Following \cite{Freidel:2010aq,Freidel:2010bw,Haggard:2012pm,Freidel:2013fia}, we shall refer to these `coherent' 4-simplices as twisted.

There has been great interest in understanding and interpreting what type of degrees of freedom are added by `twisting', (see \emph{e.g.} \cite{Freidel:2013bfa,Freidel:2013fia}) and what is their fate in the continuum limit, which could potentially have important phenomenological implications \cite{Dittrich:2021kzs}.

Our key observation, developed in \S\ref{sec:SAM}, is that these degrees of freedom are encoded in an \emph{area-metric}, which instead of defining the inner product of vectors as happens with the more traditional length-metric, determines the inner product of bi-vectors. We see a great potential in this new perspective to provide new insights into spin foam geometries and their continuum dynamics: one can now study effective actions based on the area-metrics to model spin foam dynamics.  Recent work  found that area-metrics  can be used to describe the dynamics of spin foams in the continuum limit  \cite{Dittrich:2021kzs,Dittrich:2022yoo,Borissova:2022clg}. Here we provide the underlying microscopic match of spin foam degrees of freedom to the area-metric.

More generally this is also a framework that potentially generalizes Regge calculus \cite{Regge:1961px} from length to area-metrics (and does not necessarily agree with area-Regge calculus \cite{Barrett:1997tx,Asante:2018wqy}, but could lead to a related effective dynamics).

The match is provided as follows: just as for any simplex one can reconstruct a simplicial length-metric adapted to its edge vectors from knowing their lengths, we show how to construct an area-metric for a 4-simplex given the above-mentioned 20 parameters, which may be the ten triangle areas and ten 3D-dihedral angles (corresponding to the intertwiners in LQG jargon), two at non-opposite edges per tetrahedron. Remarkably, area-metric geometries have also recently appeared in the continuum limit of area-Regge calculus\footnote{Note that area-Regge calculus associates only 10 degrees of freedom to a 4-simplex, which are given by the areas of its 10 triangles. The reconstruction of an area-metric in \cite{Dittrich:2022yoo} relies not on the data associated to a single simplex, but rather on the data associated to a regular hyper-cubic lattice.} \cite{Barrett:1997tx,Asante:2018wqy}, which is closely related to the perturbative continuum limit of spin foams \cite{Dittrich:2021kzs,Dittrich:2022yoo}. Recent work has also constructed a candidate effective action for the continuum limit of spin foams, which is based on area-metrics \cite{Borissova:2022clg}.

After we establish this result, we will proceed to define the notion of signature and triangle inequalities for twisted geometries in \S\ref{ssec:realizability}. We emphasize in \S\ref{ssec:realizability} that these generalized triangle inequalities could provide new ways to understand relative suppressions and causal aspects of spin foams. In \S\ref{ssec:pt} we first establish how to glue area-metric simplices and then discuss how to construct a notion of parallel transport. We will see that the definition of a unique parallel transport requires the specification of additional structure or additional conditions, but leave the exploration of such options to future work.

But first we will discuss more in detail aspects of classical and quantum simplicial geometry, which we now turn to do.

\subsection{(Quantum) simplicial geometry \label{ssec:QSG}}

As intuition suggests the full geometry of a tetrahedron can be specified by its edge lengths. However, other parametrizations are possible and even desirable. In $(3+1)$D LQG the area operator appears as a more fundamental (and simple) observable \cite{Rovelli:1994ge,Ashtekar:1996eg}, whereas there are different versions of length operators, which are all composite \cite{Thiemann:1996at,Bianchi:2008es}. Furthermore, the quantum labels of a spin network state can be interpreted in terms of areas and 3D dihedral angles associated to an underlying triangulation.

An alternative parametrization of simplicial geometries is provided by Minkowski's theorem, which states that a collection of $n$ vectors $\{\vec p_m\}_{m=1}^N\subset \mathbb R^{p,q},\quad p=0,1$ that \emph{closes}, that is satisfies
\begin{equation}
    \sum_{m=1}^N \vec p_m=\vec0,
    \label{eq:closure}
\end{equation}
can be used to reconstruct a unique \emph{flat} convex polytope with $n$ co-dimension-one convex polytopes having normals $\vec p_m$ and corresponding co-dimension-one volumes $p_m:=\sqrt{\vec p_m^2}$.

In particular, note that in three dimensional space, given four areas $p_m$ and two products of normals $p_{12}=\vec p_1\cdot \vec p_2$, $p_{13}=\vec p_1\cdot \vec p_3$ we may make use of the closure condition and isometry group to reconstruct the normals and thus a tetrahedron (consider \emph{e.g.} the inner product of a normal with the closure condition). Observe also that because we know the areas, the new information provided by these products are 3D-dihedral angles of two non-opposite edges (the sub-index 1 is repeated).

The only obstruction to such reconstruction is a reality condition on the solutions of the resulting system of equations. If there is no real solution, it means that there can be no tetrahedron with those areas and normal products and we say that the geometric data is not \emph{realizable} into a tetrahedron: this simply gives us a higher dimensional (and signature dependent) notion of triangle inequalities.

One can then quantize this geometry by associating operators $\hat{\vec p}_i$ to the normals. Moreover, a canonical analysis of general relativity formulated with the Holst action suggests that the upgrade should be done with the $\frak{su}(2)$ generators $\hat{\vec J}$, or their time-like and null analogues. More precisely let us consider the space-like case, where we have $\hat{\vec p}_i=\gamma\hat{\vec J} _i$ in Planck units, with $\gamma$ the dimensionless Barbero-Immirzi parameter.

The resulting Hilbert space of quantum geometry is then $\mathcal H=\bigoplus_{j_i}\text{Inv}\bigotimes_{i=1}^4 \mathcal H_{j_i}$, where $\mathcal H_{j_i}$ is the representation space of the $SU(2)$ irreducible representation of spin $j_i$. The Inv indicates that we consider the subspace of states invariant under special orthogonal transformations (as we should for geometry), so it is simply the kernel of the generator of overall `rotations' $\sum_i\hat{\vec J}_i$, also known as Gau\ss~constraint. Note therefore that we have a remarkable correspondence between orthogonal invariance and the closure relation \cite{Rovelli:2014ssa}.

With this quantization of the normals we see that the area squared operators amount to the Casimir of the $SU(2)$ representations and commute with each other. But this is not the case for $\hat p_{12}$ and $\hat p_{13}$, as \cite{Asante:2021zzh}
$$[\hat p_{12},\hat p_{13}]=\pm \frac 92 \gamma \widehat{\text{Volume}}\text{(Tetrahedron)}.$$
In fact, it can be shown that the four area operators and one of the $\hat p_{mn}$ provide a complete set of commuting observables, so that their eigen-states provide a basis for $\mathcal H$. Therefore, their  values give us an example of five quantum numbers defining the quantum geometry of a tetrahedron (\emph{cf. \S\ref{sec:intro}}). Remarkably, this quantum geometry is `fuzzy': we cannot measure the missing datum to define its classical geometry within this eigen-basis, in fact the missing datum, say $p_{ik}$ has maximal uncertainty for any element of this basis.

It is also because of this non-commutativity that when gluing two tetrahedra together, the geometry of their shared triangle $t$ cannot be matched completely \cite{Dittrich:2008ar,Dittrich:2010ey}: We can begin by fixing the areas $p_{\tau_i}$ of the triangle as seen from the tetrahedra $\tau_i$ to be equal. In this way, any remaining non-matching lays in the shape of the triangles. Therefore, it is natural to try and match the 2D angles $\alpha^k_{2D}$ as computed from each $\tau_i$. In this way, we end up with the constraints \cite{Dittrich:2008va}
$$p^{\tau_1}_t-p^{\tau_2}_t\overset{!}{=}0\quad\text{and}\quad \alpha^k_{2D}(\vec{p}^{\tau_1}_{m})-\alpha^k_{2D}(\vec{p}^{\tau_2}_n)\overset{!}{=}0,\, k=1,2.$$
However, quantum mechanically we can only implement the first set of constraints, as the functional form of $\alpha_{2D}$ involves \emph{several} products $p^{\tau_l}_{mn}$ such that the spin network states are not annihilated, precisely because they do not diagonalize more than one $p^{\tau_k}_{mn}$. In this sense, we say that the triangles are not \emph{shape-matched}. Also, one can compute the commutator between the constraints in the second set \cite{Dittrich:2008ar,Dittrich:2010ey,Dittrich:2012rj}, which shows that these are non-vanishing, and more specifically second class. That is the full set cannot be implemented sharply into the quantum theory.  Thus, as foretold, when gluing five tetrahedra to form a quantum 4-simplex, we obtain a geometry given by $5\times5-10=15$ quantum numbers. 

We can then consider that the way to resemble classical geometry as much as possible is one that minimizes the uncertainty of $p_{12}$ and $p_{13}$, \emph{i.e.} a coherent state. And such states, \emph{e.g.} \cite{Calcinari:2020bft,Freidel:2013fia}, will therefore define a semi-classical 4-simplex with $10+2\times5=20$ degrees of freedom: ten products of triangle normals, two per each tetrahedron; and ten areas. As we will now see, this over-parametrization of a 4-simplex classical geometry has a neat interpretation in terms of area-metric geometries. 

\section{Area-metric simplex\label{sec:SAM}}
\subsection{Summary of area-metrics}

Let us therefore define what an area-metric geometry is following the seminal works \cite{Schuller:2005ru,Schuller:2005yt,Punzi:2006nx}: It is a manifold $\mathcal M$ equipped with a metric $G(p)$ for $\bigwedge^2 T_p \mathcal M\quad\forall p\in\mathcal M$, and a smoothness condition for $p\rightarrow G(p)$. That is, $G$ is a rank $\binom{0}{4}$ tensor such that $G_{abcd}=G_{cdab}$ (symmetry of inner product) $G_{abcd}=-G_{bacd}(=-G_{abdc})$ —anti-symmetry of wedge product— and such that the linear map
$$G:{\bigwedge}^2 T_pM\rightarrow\left({\bigwedge}^2 T_pM\right)^*,\quad (A^{ab})\rightarrow (G_{abcd}A^{cd})$$
is invertible. This definition is justified because \emph{simple} bi-vectors $A\in \bigwedge^2 TM$, \emph{i.e.} a bi-vectors satisfying $A=l\wedge r$, are naturally interpreted as directed areas. Indeed, note that in a three dimensional flat (sub-)space $\vec l\wedge\vec r=*(\vec l\times\vec r)$, with $*$ representing the Hodge dual.

Therefore, we consider that the inner product defined by $G$ quantifies the `size' of simple bi-vectors $A=l\wedge r$ as well as angles between pairs of them.\footnote{Compare with the Euclidean formula $a\cdot_E b=|a||b|\cos\angle_a^b$ or its Lorentzian analogues \cite{Sorkin:2019llw,Asante:2021zzh}.} The former are naturally identified as the area of the parallelograms defined by the vectors $l$ and $r$, and in the case the pairs of bi-vectors span planes that intersect in a line, the latter can be interpreted as 3D dihedral angles.

As could be expected, any (pseudo-)Riemannian length-metric $g$ induces an area-metric
\begin{equation}
    G_g=(G_{abcd})=(g_{ac}g_{bd}-g_{ad}g_{bc}),
    \label{eq:induced_G}
\end{equation}
but certainly not all area-metrics are of this form as a simple counting argument confirms: In four dimensions, the index structure of $G$ gives us 21 degrees of freedom, but length-metrics have only 10. One should therefore not expect a unique procedure to extract a length-metric from an area-metric. In fact, we are aware of two different proposals to do so: The proposal in \cite{Punzi:2006nx} is motivated from properties of light propagation. The more recent work \cite{Borissova:2022clg} adapts a parametrization of bi-vector fields  in terms of Urbantke metrics and additional fields \cite{Freidel:2008ku,Speziale:2010cf}, in order to define a length-metric from an area-metric.  These techniques appeared in the analysis of modified Plebanski theories \cite{Krasnov:2009iy}, and lead to candidate effective actions for the continuum limit of spin foams \cite{Borissova:2022clg}.

The 21 degrees of freedom of a general area-metric are in mismatch with the 20 degrees of freedom of a twisted simplex. We can however restrict to the \emph{cyclic} sub-class of area-metrics, \emph{i.e.} those satisfying\footnote{\label{fnote:pseudo}In principle one can loosen the cyclicity requirement and set the left-hand side to a function that allows us to do the inversion referred to below equation \eqref{eq:4Dangle2}.}
$$G_{0123}+G_{0231}+G_{0312}=0.$$
This constraint therefore results in area-metrics with twenty degrees of freedom.

There are different motivations to introduce this restriction: Area-metrics, which are induced by a length-metric are cyclic. The restriction to cyclic area-metrics appears also in the framework of modified Plebanski theories, which are proposed to model the continuum limit of spin foams \cite{Borissova:2022clg}.  Furthermore, in a universe in which only areas can be measured, all measurable geometry can be encoded in a cyclic area-metric \cite{Borissova:2022clg}. The one acyclic component of the area-metric and the remaining 20 cyclic components span different representations of $SL(4,\mathbb R)$ \cite{Schuller:2005yt}. Note finally, that the cyclicity condition is analogous to the algebraic Bianchi identity for the Riemann tensor. In fact, cyclic area-metrics have the same (algebraic) symmetries as the Riemann tensor.

In this paper we will mainly have in mind Lorentzian or Euclidean space-times, but leave the discussion general. We therefore must also define such notions for area-metrics and for this purpose we remark that it can be shown \cite{Schuller:2005yt} that if a length-metric has signature $(p,q)$ with $D=p+q$, the signature of its induced area-metric is
\begin{equation}(P,Q)=\left(pq,\frac{(p+q)(p+q-1)}2-pq\right).
\label{eq:signature}
\end{equation}
This signature is understood in the $6$-dimensional space of bi-vectors, that is with the bi-linear map
$$G:{\bigwedge}^2 TM\times{\bigwedge}^2 TM\rightarrow \mathbb R,\quad (U,V)\rightarrow U^IG_{IJ}V^J,$$ 
where we introduced the anti-symmetrized indexes 
$$I=(ab),\quad a<b\quad\text{and}\quad J=(cd),\quad c<d$$
for the $\left(\frac{D(D-1)}2\right)$-dimensional bi-vector spaces. From this follows that $g$ is Lorentzian, \emph{i.e.} $p$ or $q$ equal $D-1$, if and only if $P=D-1$. And likewise, $g$ is Euclidean iff $P=0$. Therefore, in general we will say that an area-metric is Euclidean (Lorentzian) if $P=0$ ($P=D-1$)
\cite{Schuller:2005yt}.

\subsection{Area-metric of a (twisted) simplex}\label{Areatwisted}

In any dimension, given a set of lengths that \emph{realize}\footnote{See the discussion below eq. \eqref{eq:closure}.} a flat simplex, one can construct a matrix representation of the standard flat metric with respect to the basis defined by a corner of the simplex. Further, if one uses this algorithm then realizability is equivalent to asking the resulting matrix to have the proper signature. \cite{Sorkin:1975ah}

We will now construct an analogue of this algorithm for twisted 4-simplices and cyclic area-metrics. That is, given our ten $p_m$ and ten $p_{mn}$ (two per tetrahedron), we will reconstruct an array of $G$-components $G(a,b;c,d)$ that can be interpreted as a matrix representation of an area-metric with respect to a given basis, \emph{i.e.}
$$G(a,b;c,d)\sim \langle a\wedge b,c\wedge d\rangle.$$
Let us denote the vertices of our simplex as $(0),\dots,(4)$. and the vector going from vertex $(i)$ to vertex $(j)$ as $\overline{ij}$. Our basis will be the \emph{bi-vector simplicial corner} in $(0)$, \emph{i.e.} the one given by the vectors $\overline{0i}$. With this, a complete set of  entries of $G$ is
{\scriptsize
\begin{gather*}
    \mathcal C=\left\{\Gz{i}{j}{k}{l}|i<j,k<l\text{ \& }i+j\le k+l;i,j,k=1,\dots,4\right\}\\
    =\\
    \bigl\{\Gz{1}{2}{1}{2},\Gz{1}{2}{1}{3},\dots,\Gz{1}{2}{3}{4},\\
    \Gz{1}{3}{1}{3},\dots,\Gz{1}{3}{3}{4},\\
    \dots,\\
    \Gz{2}{3}{2}{3},\dots,\Gz{2}{3}{3}{4},\\
    \dots,\\
    \Gz{3}{4}{3}{4}\bigl\}.
\end{gather*}
}
This set of components is independent if we remove one of the entries with all vectors different, as per the cyclicity condition.

There are two main types of elements in $\mathcal C$, those with repeated vectors and those without, let us first reconstruct the former. They are all related to the geometry of the four tetrahedra that have $(0)$ as a vertex and after possible re-arrangements of the vectors also come in two forms: $\Gz{i}{j}{i}{j}$ and $\Gz{i}{j}{i}{k}$. Of the first kind we have six and they are clearly identified with six of our $p_m$ variables by
$$\Gz{i}{j}{i}{j}=(2\text{Area}(\triangle_{(0ij)}))^2,$$
where in the right hand side we have the area of the triangle with vertices $(0)$, $(i)$ and $(j)$, a $p_m$ variable.

There are 12 independent matrix elements $\Gz{i}{j}{i}{k}$ of the second form.  To each tetrahedron of type $(0ijk)$, \emph{i.e.} to each of the four tetrahedra sharing the vertex $0$, we can associate three of these matrix elements of second kind. Of these three we can directly identify two with the two $p_{mn}$ variables per tetrahedron. To match the remaining third elements, we use that for an arbitrary tetrahedron $(oijk)$ the following relation 
\begin{equation}
    \overline{oi}\wedge\overline{oj}+\overline{oj}\wedge\overline{ok}+\overline{ok}\wedge\overline{oi}+\overline{ik}\wedge\overline{ij}=0
    \label{eq:closurep}
\end{equation}
holds.
This can be shown by writing $\overline{ik}=\overline{io}+\overline{ok}$, and similarly for $\overline{ij}$; expanding and using the properties of $\wedge$. Eq. (\ref{eq:closurep})  is essentially the closure relation —\emph{cf.} eq. \eqref{eq:closure}. We can use this relation to compute the remaining entries of the form $\Gz{i}{j}{i}{k}$ by taking inner products with \eqref{eq:closurep}. This will require us to employ the geometric data of the tetraheda $(0ijk)$ we have not used so far, \emph{i.e.} the remaining four areas $p_m$.

For example, let us focus on the tetrahedron $(0ijk)=(0123)$ and identify the normals as $\vec p_1\sim\overline{01}\wedge\overline{02}$, $\vec p_2\sim\overline{03}\wedge\overline{01}$, $\vec p_3\sim \overline{02}\wedge\overline{03}$ and $\vec p_4\sim\overline{13}\wedge\overline{12}$. Now let us suppose that $p_{12}$ and $p_{13}$ are directly matched with $\Gz{1}{2}{3}{1}$ and $\Gz{1}{2}{2}{3}$, respectively. Then we can use \eqref{eq:closurep} to determine $p_{23}\sim\Gz{3}{1}{2}{3}$ by taking the equation's inner product with $\overline{13}\wedge{\overline{12}}$ to get
\begin{align*}
    0&=\G{1}{3}{1}{2}{0}{1}{0}{2}+\G{1}{3}{1}{2}{0}{2}{0}{3}\\
    &\quad+\G{1}{3}{1}{2}{0}{3}{0}{1}+\G{1}{3}{1}{2}{1}{3}{1}{2}\\
    &=p_4^2-p_1^2-p_2^2-p_3^2-2p_{12}-2p_{13}-2\Gz{3}{1}{2}{3},
\end{align*}
where in the last step we used $\overline{lm}=\overline{l0}+\overline{0m}$ and the properties of $G$. From such equation we can immediately obtain $p_{23}\sim\Gz{3}{1}{2}{3}$ in terms of our given variables. Notably, the solution depends on $p_4^2\sim\G{1}{3}{1}{2}{1}{3}{1}{2}$, which is a geometric variable we had not used previously.

In summary, we can determine all entries of the forms $\Gz{i}{j}{i}{j}$ and $\Gz{i}{j}{i}{k}$ by using the geometric data associated to the tetrahedra $(0ijk)$: ten areas $p_m$ and eight $p_{mn}$'s. Thus, if we can reconstruct two of the entries with no repeated index, using cyclicity we will have completely determined the elements of $\mathcal C$ and whence $G$. Our two remaining $p_{mn}$ variables associated to the remaining  tetrahedron $(1234)$ allow us to do so.

Without loss of generality let us assume that the  two remaining $p_{mn}$ variables are identified with $\G{1}{3}{1}{2}{1}{3}{1}{4}$ and $\G{1}{2}{1}{3}{1}{2}{1}{4}$. Then, using that $G$ is multi-linear and its symmetry properties, we see that
{
\scriptsize
\begin{align}
    \G{1}{3}{1}{2}{1}{3}{1}{4}&=G\left(\overline{10}+\overline{03},\overline{10}+\overline{02};\overline{10}+\overline{03},\overline{10}+\overline{04}\right)\nonumber\\
    &=-\Gz{1}{2}{3}{4}-\Gz{1}{4}{3}{2}\nonumber\\
     &\quad-\Gz{1}{2}{1}{3}-\Gz{1}{3}{2}{3}\nonumber\\
     &\quad-\Gz{1}{3}{1}{4}+\Gz{1}{3}{3}{4}\nonumber\\
     &\quad+\Gz{1}{2}{1}{4}-\Gz{2}{3}{3}{4}\nonumber\\
     &\quad+\Gz{1}{3}{1}{3}\nonumber\\
     &\quad=:-\Gz{1}{2}{3}{4}-\Gz{1}{4}{3}{2}\nonumber\\
     &\quad\quad+C_{3234},
     \label{eq:4Dangle1}
\end{align}
}
where we note that $C_{3234}$ includes only terms $\Gz{i}{j}{i}{k}$ of the second form, which we already expressed as functions of $p_m$ and $p_{mn}$.

Similarly, from the exchange $2\leftrightarrow3$ we see that
{
\small
\begin{align}
    \G{1}{2}{1}{3}{1}{2}{1}{4}&=-\Gz{1}{3}{2}{4}-\Gz{1}{4}{2}{3}\nonumber\\
     &\quad+C_{2324}\nonumber\\
     &=-\Gz{1}{2}{3}{4}+2\Gz{1}{4}{3}{2}\nonumber\\
     &+C_{2324},
     \label{eq:4Dangle2}
\end{align}
}
where in the last step we used cyclicity.

Equations \eqref{eq:4Dangle1} and \eqref{eq:4Dangle2} define a linear system of equations that allow us to determine $\Gz{1}{2}{3}{4}$ and $\Gz{1}{4}{3}{2}$ in terms of $C_{3234}$, $C_{2324}$, and two unused geometric parameters $p_{mn}$ which we identified with $\G{1}{3}{1}{2}{1}{3}{1}{4}$ and $\G{1}{2}{1}{3}{1}{2}{1}{4}$.

Thus, as advertised, from the geometric data of a twisted 4-simplex we can reconstruct the set of $G$-elements $\mathcal C$ and from there the a whole tensor $G$ that satisfies the linear and index properties of a cyclic area-metric, \emph{i.e.} an area-metric tensor. Reciprocally, from an area-metric tensor we can reconstruct the geometric data of a twisted 4-simplex. Thus, a twisted simplex is an area-metric simplex.

\subsection{Realizability (generalized triangle inequalities)\label{ssec:realizability}}

A remark however is in place. Above we mentioned that from \emph{realizable} length data for a simplex, one can reconstruct a simplicial length-metric with a well established signature. Yet, in the previous section we did not discuss realizability (or signature). Indeed, such a notion has not been defined for twisted geometries yet.

In a sense in spin foams realizability is not of prime importance: Some models, for example, give non-vanishing amplitudes, albeit exponentially suppressed, for configurations that violate top-dimensional length-geometric triangle inequalities. Viewed in this way, spin foams may implement top-dimensional triangle inequalities only `weakly'. This is, in a sense, another manifestation of how spin foams construct top-dimensional geometry from co-dimension one geometries.

For example, the (Euclidean) Ponzano-Regge model gives a zero amplitude for configurations that violate the (proper) triangle inequalities. However, there are non-vanishing, but exponentially suppressed amplitudes for configurations that violate the inequality for the tetrahedron \cite{Barrett:1993db}, so we have four well defined triangles which cannot be glued to form a tetrahedron in Euclidean space.\footnote{\label{fnote:HH}
However, they can form a tetrahedron in Lorentzian space \cite{Barrett:1993db}, so this could be desirable feature: it can, for example, be used for Lorentzian Hartle-Hawking no-boundary-like scenarios \cite{Hartle:1983ai,Dittrich:2021gww,Asante:2021phx}. Importantly, it is also a consequence of triangulation invariance \cite{Dittrich:2021gww}.
}

A similar situation occurs in the Lorentzian EPRL model \cite{Engle:2007wy}, where one can have non-vanishing, but exponentially suppressed amplitudes for Euclidean 4-simplices \cite{Han:2021rjo} (which violate Lorentzian 4D inequalities, but satisfy lower dimensional ones).

Thus, although not strictly, spin foam models do  care about realizability (and signature), so in this spirit we now turn to defining its area-geometric version.

As mentioned in \S\ref{sec:SAM}, for Lorentzian and Euclidean geometries, realizability of a length-simplex amounts to the simplicial length-metric having the proper signature. Indeed, this means that there is a change of basis mapping the length-reconstructed $g$ to the standard length-metric of $\mathbb R^{p,q}$ in the canonical basis. Such a map transforms the canonical vectors into $D$ linearly independent vectors that form the corner of a $D$-simplex and thereof the simplex itself, so it is indeed realizable. Likewise, we propose that a set of 20 variables $\{p_{mn},p_m\}$ \emph{realize} a twisted 4-simplex if its area-tensor $G$ has the right signature, as discussed around eq. \eqref{eq:signature}. In other words, we ask that there is a \emph{diagonalizing} bi-vector basis in which 
\begin{equation}
    (G_{AB})\overset{\cdot}{=}\text{diag}(\pm 1,\pm 1,\pm 1,\pm 1,\pm 1,\pm 1),
    \label{eq:induced_diag}
\end{equation}
with the number of positive and negative signs given by the signature.

We remark that then, the matrix representation \eqref{eq:induced_diag} agrees with the one obtained by considering the area-metric induced by the standard length-metric of $\mathbb R^{p,q}$ when written in a wedge-product basis of an orthonormal basis (\emph{i.e.} using \eqref{eq:induced_G} in an orthonormal basis and going to the antisymmetrized indices). This does {\it not} mean that our area-metric $G$ is induced by a length-metric as the diagonalizing basis may not be simple.

Note that this notion of realizability naturally gives a definition of signature for a twisted simplex, which in turn implies that in the Lorentzian context, shape-matched data associated to a Euclidean simplex would not be realizable, so that the exponential suppression of the Lorentzian EPRL/FK amplitude mentioned above (see also footnote \ref{fnote:HH}) \cite{Han:2021rjo} is not surprising from this point of view. Another class of area-geometries which is exponentially suppressed are non-shaped (twisted) data --- this holds by construction in effective spin foams \cite{Asante:2020qpa,Asante:2020iwm,Asante:2021zzh}  and has been observed to hold in the EPRL/FK model \cite{Han:2021kll,Asante:2022lnp}, so this naturally raises the question of whether there is a hierarchy of suppressed data that can be re-casted in terms of area-geometries. In fact, such study could also provide a refinement for our definition of realizability. We find the possibility of casting spin foam amplitude suppression in terms of area-geometries very interesting, but we leave it for future works.

As foreshadowed above, our proposed definition of realizability is not only a direct and minimal generalization of the length-metric condition, but offers the following picture: If a simplicial area-metric $G$ satisfies this condition, then there is a change of basis for the $\left(\frac{4(4-1)}2=6\right)$-dimensional space of bi-vectors mapping $G_{AB}$ into its representation in an orthonormal basis. Note that such a map is not unique, as we can always compose it with an orthogonal transformation. However, in complete analogy to the length-geometric case, any such map provides a notion of simplicial corner in the space of bi-vectors. From this point of view, the space of twisted simplices is equivalent to\footnote{Depending on whether one wishes to distinguish between simplices of different orientation (as done in spin foams) or not (as happens with Regge calculus), we should replace O(P,Q) with SO(P,Q).} GL$(6,\mathbb R)/$O$(P,Q)$ up to the implementation of cyclicity (this indeed gives us $6\times 6-\frac{6(6-1)}2-1=20$ degrees of freedom).

Our definition also satisfies two necessary criteria:
\begin{enumerate}
    \item It implies that all tetrahedra are realizable in the length-geometric sense (twisting comes from gluing tetrahedra, they themselves are not twisted) and
    \item When shape matching is satisfied, it reduces to length realizability.
\end{enumerate}
Let us prove so. 
\begin{enumerate}
    \item holds because: From the reasoning above, we can find a transformation that sends an orthonormal bi-vector basis $\{E_I\}$, to a bi-vector simplicial corner that we may choose such that it includes the bi-vectors of any tetrahedron in question. Because $G_{AB}$ is real and symmetric, the map can be of the form $U^{-1}=U^T$ and can be followed by an orthogonal transformation $O$ that we can pick so that the tetrahedron's bi-vectors are expressed as a linear combination of just three $E_I$ bi-vectors.
    
    This defines a three dimensional linear transformation $T$ that sends the three basis bi-vectors into the tetrahedral bi-vectors. The signature of the former, which we will now denote as $u_\mu$, $\mu=1,2,3$; determines the signature of the putative tetrahedron: let it be $(\pi,\text\textrevglotstop)$ —with $\pi+\text\textrevglotstop=3$. The identification $u_\mu\sim\vec e_\mu$ with the canonical basis $\vec e_\mu$ of $\mathbb R^{\pi,\text\textrevglotstop}$ is then natural and we expect $\vec v_\mu=T\vec e_\mu$ (defined through the linear combination above) to give the normals of our tetrahedron.
    
    Let us therefore consider the four vectors
    $$\left\{\vec v_\mu,\vec v_4=-\sum_{\mu=1}^3\vec v_\mu\right\}.$$
    These close by construction and therefore define a tetrahedron in $\mathbb R^{\pi,\text\textrevglotstop}$ by virtue of Minkowski's theorem (\emph{cf.} 
    \S\ref{ssec:QSG}), we only need to confirm that its geometry corresponds to the relevant values of $p_m$ and $p_{mn}$ from which we constructed $G_{AB}$.
    
    Let $\gamma$ be the flat metric of $\mathbb R^{\pi,\text\textrevglotstop}$, then the inner product between these normals for $\mu,\nu\le3$ is
    \begin{align*}
    p_{\mu\nu}
    &=v_\mu^\rho \gamma_{\rho\sigma}v_\nu^\sigma=(T\vec e_\mu)^\rho \gamma_{\rho\sigma}(T\vec e_\nu)^\sigma\\
    &=\sum_{I,J}{}^{'}(T u_\mu)^I \Gamma_{IJ}(T u_\nu)^J\\
    &=\sum_{I,J}(OU)^I_\mu \Gamma_{IJ} (OU)^J_\nu,
    \end{align*}
    where we introduced the metric $\Gamma$ of $\mathbb R^{P,Q}$ and the primed sum indicates that we only sum over the indices $I,J$ associated to $\mu$ and $\nu$. However, as done in the last step, we can extend the sum over the whole range $I,J=1,\dots,6$ because our construction is such that $(T u_\rho)^K$ vanishes if $K$ is not related to a Greek index.
    
    Now, the latter expression can be rewritten as $\left((OU)^T\Gamma (OU)\right)_{\mu\nu}$, which is $G_{\mu\nu}$ by construction! So
    $$p_{\mu\nu}=G_{\mu\nu}$$
    with the proper identification of indices in $G$.
    
    Thus, any potential difference between $p_{mn}$ and $p_{\mu\nu}$ (or $p_m$ and $p_\nu:=p_{\nu\nu}$) has to come from the products with $\vec v_4$. But since these can be expressed in terms of the above $p_{\mu\nu}$'s with $\mu,\nu<4$ due to closure, and since $G_{AB}$ is consistent with this closure by the construction in section \ref{Areatwisted}, there is no difference at all.
    
    Therefore, we have found four vectors that close and whose geometry is consistent with our tetrahedral data, making the latter realizable.
    
    Thus, 1. indeed holds and therefore also does
    
    \item because now that we know that our three dimensional sub-simplices are realizable, so are all proper sub-simplices and therefore we only need to prove that when shape matching is satisfied, the formally defined volume squared of the simplex has the correct sign associated with its signature. Indeed, this is another characterization of realizability, namely: all proper sub-simplices are realizable and the formal volume squared has the right signature-dependent sign. \cite{Asante:2021zzh,Tate:2011rm}.
    
    Now, if shape matching is satisfied then $G_{AB}$ can be induced from a simplicial length-metric $g_{\mu\nu}$ which is obtained by writing the lengths in terms of areas. This is possible because a 4-simplex has the same number of triangles as edges and the length-area system is invertible locally in configuration space.\footnote{Such system may have several roots but the roots can be determined by also matching the values of the 3D dihedral angles.} Then it can be shown that \cite{Punzi:2006nx}
    $$\text{det}(G_{AB})=\left(\text{det}g\right)^{D-1}$$
    and therefore in our case with $D=4$ the sign of these determinants is equal. But the latter provides the formal definition of the volume squared of the simplex (up to a positive factor) and therefore, if $G_{AB}$ is realizable in our sense the volume squared has the right sign, making our 4-simplex realizable when shape matching is satisfied.
\end{enumerate}

Therefore it is indeed the case that our definition of realizability not only minimally generalizes its length-geometric counterpart, but satisfies two tests that any definition should. However, we leave open the possibility for refinements of this definition, \emph{e.g.} parallel transport considerations (see discussion below) suggest that simplicity may also need to be considered.

\section{Gluing area-metric simplices}\label{ssec:pt}

So far we have shown that for a 4-simplex with twisted (boundary) data, we have a well defined area-metric for it that determines all of its geometry and vice-versa. We defined a notion of realizability, that ensures that the area-metric has a signature consistent with Euclidean or Lorentzian space-time respectively.

Twisted simplices can be also glued to each other: for a given pair of to be glued 4-simplices the data associated to the to be shared tetrahedron, that is the 6 $p$-variables, have to match. This gluing is inherited by the area-metric 4-simplices. Given the area-metrics associated to a pair of to be glued 4-simplices, we can compute the 20 $p$-parameters associated to each simplex, and check whether the 6 $p$-parameters associated to the to be shared tetrahedron match. Additionally, we demand that realizability holds for both simplices with respect to the same space-time signature.

We can thus glue area-metric simplices to a triangulation that can be considered as a discretization of an area-metric space-time. In order to construct geometric quantities such as a generalized curvature, it would be helpful to have a notion of parallel transport. In the following we will sketch what has to be considered to construct a unique notion of parallel transport, but leave this for future work. We note that the issue of area-metric parallel transport has also not been fully resolved in the continuum:  The construction of a parallel transport in \cite{Schuller:2005ru} requires additional structure,  \emph{e.g.} an additional length-metric. One can, associate a length-metric to a given area-metric: \cite{Schuller:2005ru} and \cite{Borissova:2022clg} propose two different procedures to do so, the construction in \cite{Borissova:2022clg} is inspired by the modified Plebanski formalism, and therefore nearer to spin foams, but requires additional work to translate into the discrete, which goes beyond the current paper.

To start with, we first recall how to construct a parallel transport for length-metric simplices in any dimension $d$. Such a parallel transport can be defined between two neighbouring $d$-simplices $\sigma_1$ and $\sigma_2$. To this end we assume that the length-metrics $g_1$ and $g_2$ are given with respect to the $d$ basis vectors of a simplicial corner for $\sigma_1$ and the $d$ basis vectors of a simplicial corner for $\sigma_2$, respectively. The simplicial corners should be defined at a vertex that is shared by both $d$-simplices, so that the two bases coincide in $(d-1)$ of their vectors. (The existence of a pair of such bases is guaranteed by the notion of realizability for length-metric simplices and by the gluing conditions for such simplices, which enforce that the length geometries of the shared $(d-1)$-simplex coincide.)

The parallel transport matrix $U$ transforms the first basis into the second basis, keeping the shared $(d-1)$ vectors invariant. Thus $U$ also transforms $g_1$ into $g_2$, that is
\begin{equation}\label{equ:PT1}
U^T g_1 U\,=\, g_2    \, .
\end{equation}
Given $g_1$ and $g_2$ we can use (\ref{equ:PT1}) to solve for the matrix elements of $U$. Note, that the conditions above leave only $d$ unknown matrix elements in the $d\times d$ matrix $U$. These conditions also ensure that the symmetric matrices $g_1$ and $g_2$ agree in a $(d-1)$-dimensional sub-matrix. We are thus left with $d$ equations, which can be used to specify the $d$ unknown variables, and therefore the parallel transport matrix $U$.

Let us now apply the same concept to area-metric simplices in four dimensions. Here we again assume that the area-metrics $G_1$ and $G_2$ are expressed with respect to a bi-vector basis arising from a simplicial corner for $\sigma_1$, and a bi-vector basis arising from a simplicial corner for $\sigma_2$, respectively. The two simplicial corners are attached to a shared vertex, hence {\it three} of the six bi-vectors in each of the two bases agree, and the parallel transport matrix $U$ is constrained to fix these three vectors. To determine $U$ we have then to solve 
\begin{equation}\label{equ:PT2}
U^T G_1 U \,=\, G_2 \, .
\end{equation}
This time we have 21 equations, of which 6 are automatically satisfied on account of the gluing conditions. $U$ has 36 elements of which $6\times 3$ are already determined by the condition that the two bases agree in three of their bi-vectors. We thus have only 15 equations for 18 unknowns and are left with a 3-parameter ambiguity.

This can be also understood as follows: Realizability implies that we can find simplicial corners for both $\sigma_1$ and $\sigma_2$. We would like to rotate the latter so that the bi-vectors of the to-be-shared tetrahedra match and we can certainly find such orthogonal transformation, but it is not unique: once we have found one, it can be followed by another arbitrary rotation that fixes the three bi-vectors associated to the tetrahedral corner. The dimension of this subgroup is $\binom{3}{2}=3$.

This discussion shows that the problem with parallel transport is not with existence but with uniqueness. Thus it is possible for two glued area-metric simplices to find a common bi-vector basis, so that the induced area-metric is constant across both 4-simplices. This is a property shared by length-metric simplices, where pairs of neighbouring simplices admit a common vector basis.

The issue with uniqueness results from working with bi-vector bases. We were forced to consider bi-vector bases, as we needed to define a standard area-metric (for a given space-time signature), to which all (realizable) area-metrics can be transformed. This appeared first  for our notion of realizability. Similarly, the notion of simplicial parallel transport (in the length and area-metric case) does require a standard metric, to which the metrics associated to two neighbouring simplices can be transformed. 

The uniqueness of parallel transport might however be ensured, if we demand additional simplicity constraints for the parallel transport. In this case one has however also to ensure the existence of (real) solutions to the parallel transport equations (\ref{equ:PT2}) and the additional constraints. We leave this to future work, but note that the need to enforce simplicity conditions on the connection was also noted in the continuum framework of \cite{Schuller:2005ru}. In the future, it will also be interesting to explore the definition of a deficit angle, which could partially be constructed from these parallel transport transformations, but must have information about winding around a bone.\footnote{This might be particularly important in Lorentzian path integral approaches to Quantum Gravity based on Regge-like calculus, where histories with non-trivial windings can be exponentially suppressed or enhanced \cite{Sorkin:2019llw,Asante:2021phx,Asante:2021zzh}.}

\section{Discussion \label{sec:conclusions}}

Spin foams and Loop Quantum Gravity feature an extended geometrical configuration space \cite{Dittrich:2008va,Dittrich:2008ar,Freidel:2010aq}. The geometrical interpretation of these additional degrees of freedom remained however an open issue. In this work we propose that gravitational spin foams can be understood as discretized area-metric geometries. To this end we matched the (coherent state) parameters associated to a single spin foam 4-simplex to a (constant) area-metric geometry. We also specified how to glue such simplices to form an area-metric triangulation. We  specified realizibility conditions, that ensure that area-metric simplices have a consistent signature. Such conditions have not been formulated before for `twisted' simplices, but are now available due to the identification of twisted simplices with area-metrics.

Our reinterpretation of spin foam geometries as area-metric geometries is supported by recent insights into the continuum limit of gravitational spin foams \cite{Dittrich:2021kzs,Dittrich:2022yoo,Borissova:2022clg}, which indicates that it can be understood in terms of an {\it effective} area-metric dynamics. In this work we provided a {\it microscopic} explanation for this appearance of the area-metric, further strengthening this connection between spin foam dynamics and area-metric dynamics.

In contrast to earlier suggestions for the geometric interpretation of twisted geometries, \emph{e.g.} \cite{Freidel:2013bfa}, the link to area-metrics is therefore supported and informed by the spin foam dynamics.

There are  already various actions that can be applied to area-metric triangulations: Firstly area-angle Regge calculus \cite{Dittrich:2008va}, which was proposed as a classical description for the EPRL/FK spin foam models. The work \cite{Dittrich:2008va} was the first to specify the shape matching constraints, which reduce area-angle (and therefore area-metric) configurations to length-Regge configurations. Such shape matching conditions are missing from area Regge calculus \cite{Barrett:1997tx,Asante:2018wqy}, which can also be adapted\footnote{\emph{E.g.} one can define that the action only depends on the areas and not on the 3D angular parameters.} to area-metric simplices. More generally, the action used for the construction of effective spin foams \cite{Asante:2020qpa,Asante:2021zzh,Asante:2020iwm} interpolates between these two cases, by imposing the shape matching constraints only weakly, and can be also applied to area-metric triangulations. Finally, using coherent states, the $SU(2)$-BF action, describing a topological theory, can be also expressed in terms of the 20 parameters of the twisted/area-metric simplices \cite{Freidel:2013fia}. This is again an action where the shape matching conditions are not imposed, but differently from the area Regge action it does have a non-trivial dependence on the angular parameters. 

{
The extension of the geometrical configuration space from length to area metrics leads to  the so-called flatness problem for spin foams \cite{Bonzom:2009hw,Hellmann:2013gva,Han:2013ina,Oliveira:2017osu,Engle:2020ffj} : The classical action principle, namely the Plebanski formalism, on which spin foams are based, demands that one has to impose constraints that reduce the area metric configuration space to a length configuration space. However, due to an anomaly in the constraint algebra \cite{Dittrich:2012rj}, spin foams do impose these constraints only weakly \cite{Engle:2007uq,Freidel:2007py,Asante:2020qpa,Asante:2021zzh}, e.g. via Gaussians or coherent states, allowing fluctuations within the area metric configuration space. The width of the Gaussians imposing the constraints scales with $\sqrt{\hbar}$, whereas the frequency of the oscillations grow with $1/\hbar$. Thus in a naive classical limit $\hbar\rightarrow 0$ the oscillations win over the Gaussians, and this seems to inhibit the imposition of the constraints \cite{Asante:2020qpa}. There are two different arguments, which avoid this conclusion: Firstly, the Gaussians' width scales with the Barbero-Immirzi parameter. Indeed, in \cite{Asante:2020iwm}  (effective) spin foam expectation values were computed, and these reproduced a gravitational dynamics, if the Barbero-Immirzi parameter was chosen to be sufficiently small. Secondly, one can perform a perturbative continuum limit for effective spin foams \cite{Dittrich:2021kzs,Dittrich:2022yoo}, and show that the additional degrees of freedom which distinguish an area metric from a length metric are massive. The continuum limit is dominated by the massless degrees of freedom, which turn out to come only from the graviton \cite{Dittrich:2021kzs,Dittrich:2022yoo}.
}

Let us also note that the richness of discrete actions for area-metric triangulations is mirrored by actions available in the continuum. Apart from the earlier area-metric works \cite{Schuller:2005ru,Schuller:2005yt,Punzi:2006nx}, the recent work \cite{Borissova:2022clg} builds on the modified Plebanski action framework \cite{Krasnov:2009iy,Freidel:2008ku,Speziale:2010cf} to construct a family of continuum actions for area-metrics. The area-metrics can be split into a length-metric part and additional `pure' area-metric degrees of freedom \cite{Borissova:2022clg}. Integrating out the latter leads to an Einstein-Hilbert term and a Weyl curvature squared term. This Weyl curvature squared term can be understood as a correction, resulting from the additional area-metric degrees of freedom \cite{Borissova:2022clg}. This provides an example of how the understanding of the fundamental underlying degrees of freedom of a theory can lead to deeper insights into its dynamics. This holds in particular when working with the concept of effective actions and renormalization, which, given the fundamental variables, include \emph{a priori} all possible terms allowed by the symmetries.

We are therefore very hopeful that the notion of area-metric triangulations helps to facilitate the connection between various proposals for discrete actions, and particularly for spin foam actions, with the continuum actions.  Having a match between geometric concepts in the discrete and the continuum will also be helpful for establishing a renormalization flow for spin foams \cite{Dittrich:2014ala,Asante:2022dnj}, that can potentially connect to the continuum, and in particular to the framework of modified Plebanski theories \cite{Krasnov:2009iy}. This will give us long awaited insights into spin foam dynamics, and help to establish and understand the continuum limit and the phenomenology of spin foam dynamics.

\section*{Acknowledgements}

JPA thanks Seth Asante, Johanna Borissova, Fedro Guillén, Harish Murali and Jinmin Yi for very useful discussions. JPA is supported by a NSERC grant awarded to BD.
Research at Perimeter Institute is supported in part by the Government of Canada through the Department of Innovation, Science and Economic Development Canada and by the Province of Ontario through the Ministry of Colleges and Universities.

\bibliography{references}

\end{document}